\documentstyle[times,psfig,color,mathrsfs]{mn}

\newif\ifAMStwofonts
\AMStwofontstrue

\ifoldfss

  \newcommand{\tbd}[1] {\textbf{\color{red}{To be done later}}}
  \ifCUPmtlplainloaded \else
    \NewTextAlphabet{textbfit} {cmbxti10} {}
    \NewTextAlphabet{textbfss} {cmssbx10} {}
    \NewMathAlphabet{mathbfit} {cmbxti10} {} 
    \NewMathAlphabet{mathbfss} {cmssbx10} {} 
  \fi
  \ifAMStwofonts
    \ifCUPmtlplainloaded \else
      \NewSymbolFont{upmath} {eurm10}
      \NewSymbolFont{AMSa} {msam10}
      \NewMathSymbol{\upi}     {0}{upmath}{19}
      \NewMathSymbol{\umu}     {0}{upmath}{16}
      \NewMathSymbol{\upartial}{0}{upmath}{40}
      \NewMathSymbol{\leqslant}{3}{AMSa}{36}
      \NewMathSymbol{\geqslant}{3}{AMSa}{3E}

       \let\le=\leqslant
       \let\ge=\geqslant
    \fi
  \fi
\fi 

\ifnfssone
  \newmathalphabet{\mathit}
  \addtoversion{normal}{\mathit}{cmr}{m}{it}
  \addtoversion{bold}{\mathit}{cmr}{bx}{it}

  \newcommand{\tbd}[1] {\textbf{\color{red}{To be done later}}}
  \newmathalphabet{\mathbfit} 
  \addtoversion{normal}{\mathbfit}{cmr}{bx}{it}
  \addtoversion{bold}{\mathbfit}{cmr}{bx}{it}
  \newmathalphabet{\mathbfss} 
  \addtoversion{normal}{\mathbfss}{cmss}{bx}{n}
  \addtoversion{bold}{\mathbfss}{cmss}{bx}{n}
  \ifAMStwofonts
    \ifCUPmtlplainloaded \else
      \UseAMStwoboldmath
      \makeatletter
      \new@mathgroup\upmath@group
      \define@mathgroup\mv@normal\upmath@group{eur}{m}{n}
      \define@mathgroup\mv@bold\upmath@group{eur}{b}{n}
      \edef\UPM{\hexnumber\upmath@group}
      \new@mathgroup\amsa@group
      \define@mathgroup\mv@normal\amsa@group{msa}{m}{n}
      \define@mathgroup\mv@bold\amsa@group{msa}{m}{n}
      \edef\AMSa{\hexnumber\amsa@group}
      \makeatother
      \mathchardef\upi="0\UPM19
      \mathchardef\umu="0\UPM16
      \mathchardef\upartial="0\UPM40
      \mathchardef\leqslant="3\AMSa36
      \mathchardef\geqslant="3\AMSa3E

       \let\le=\leqslant
       \let\ge=\geqslant
    \fi
  \fi
\fi 

\ifnfsstwo

  \newcommand{\tbd}[1] {\textbf{\color{red}{To be done later}}}
  \DeclareMathAlphabet{\mathbfit}{OT1}{cmr}{bx}{it}
  \SetMathAlphabet\mathbfit{bold}{OT1}{cmr}{bx}{it}
  \DeclareMathAlphabet{\mathbfss}{OT1}{cmss}{bx}{n}
  \SetMathAlphabet\mathbfss{bold}{OT1}{cmss}{bx}{n}
  \ifAMStwofonts
    \ifCUPmtlplainloaded \else
      \DeclareSymbolFont{UPM}{U}{eur}{m}{n}
      \SetSymbolFont{UPM}{bold}{U}{eur}{b}{n}
      \DeclareSymbolFont{AMSa}{U}{msa}{m}{n}
      \DeclareMathSymbol{\upi}{0}{UPM}{"19}
      \DeclareMathSymbol{\umu}{0}{UPM}{"16}
      \DeclareMathSymbol{\upartial}{0}{UPM}{"40}
      \DeclareMathSymbol{\leqslant}{3}{AMSa}{"36}
      \DeclareMathSymbol{\geqslant}{3}{AMSa}{"3E}

       \let\le=\leqslant
       \let\ge=\geqslant
    \fi
  \fi
\fi 

\ifCUPmtlplainloaded \else
  \ifAMStwofonts \else 
    \def\upi{\pi}
    \def\umu{\mu}
    \def\upartial{\partial}
  \fi
\fi

   \title[Europium production]{Europium production: neutron star mergers versus 
     core-collapse supernovae}
   \author[F.~Matteucci et al.]{F.~Matteucci,$^{1,2,3}$\thanks{E-mail: 
       matteucc@oats.inaf.it} D.~Romano,$^{4}$ A.~Arcones$^{5,6}$, 
     O.~ Korobkin$^{7}$ and S.~Rosswog$^{7}$\\
     $^1$Dipartimento di Fisica, Sezione di Astronomia, Universit\`a di 
     Trieste, Via G.B. Tiepolo 11, I-34143 Trieste, Italy \\
     $^2$INAF, Osservatorio Astronomico di Trieste, Via G.B. Tiepolo 11, 
     I-34143 Trieste, Italy\\
     $^3$INFN, Sezione di Trieste, Via A. Valerio 2,
     I-34127 Trieste, Italy\\
     $^{4}$INAF, Osservatorio Astronomico di Bologna, Via Ranzani 1, I-40127 
     Bologna, Italy\\
     $^{5}$Institut f\"ur Kernphysik, Technische Universit\"at Darmstadt, 
     Schlossgartenstra{\ss}e 2, D-64289 Darmstadt, Germany\\
     $^{6}$GSI Helmholtzzentrum fuer Schwerionenforschung GmbH, Planckstr. 1 
     D-64291 Darmstadt, Germany\\
     $^{7}$  The Oskar Klein Centre, Department of Astronomy, AlbaNova, 
     Stockholm University, SE-106 91 Stockholm, Sweden}

\begin{document}

     \date{Accepted . Received ; in original form \today} 

     \pagerange{\pageref{firstpage}--\pageref{lastpage}} \pubyear{2013}

     \maketitle

     \label{firstpage}


   \begin{abstract}
     We have explored the Eu production in the Milky Way by means of a 
     very detailed chemical evolution model. In particular, we have assumed 
     that Eu is formed in merging neutron star (or neutron star black hole) 
     binaries  as well as in type II 
     supernovae. We have tested the effects of several important parameters 
     influencing the production of Eu during the merging of two neutron 
     stars, such as: i) the time scale of coalescence, ii) the Eu yields and 
     iii) the range of initial masses for the progenitors of the neutron 
     stars. The yields of Eu from type II supernovae are very uncertain, more 
     than those from coalescing neutron stars, so we have explored several 
     possibilities. We have compared our model results with the observed rate 
     of coalescence of neutron stars, the solar Eu abundance, the [Eu/Fe] 
     versus [Fe/H] relation in the solar vicinity and the [Eu/H] gradient 
     along the Galactic disc. Our main results can be 
     summarized as follows: i) neutron star mergers can be entirely responsible 
     for the production of Eu in the Galaxy if the coalescence time scale is no 
     longer than 1~Myr for the bulk of binary systems, the Eu yield is around 
     3~$\times$ 10$^{-7}$~M$_{\odot}$, and the mass range of progenitors 
     of neutron stars is 9--50~M$_{\odot}$; ii) both type II supernovae and 
     merging neutron stars can produce the right amount of Eu if the neutron 
     star mergers produce 2~$\times$ 10$^{-7}$~M$_{\odot}$ per system and type 
     II supernovae, with progenitors in the range 20--50~M$_{\odot}$, 
     produce yields of Eu of the order of 10$^{-8}$--10$^{-9}$~M$_{\odot}$; 
     iii) either models with only neutron stars producing Eu or mixed ones can 
     reproduce the observed Eu abundance gradient along the Galactic disc.
   \end{abstract}



   \section{Introduction}

   Approximatively half of the elements beyond the iron peak are formed through 
   rapid neutron captures in stars (r-process; Burbidge et al. 1957). `Rapid' 
   refers to the timescale of the process relative to the $\beta$-decay rates 
   of the unstable nuclei. Although detailed evaluations of the mechanisms of 
   r-process nucleosynthesis have long since been made (Burbidge et al. 1957; 
   Seeger, Fowler \& Clayton 1965), the dominant production site of the 
   r-process elements has not yet been unambiguously identified (see e.g. 
   Thielemann et al. 2010).

   The challenge identifying the main astrophysical r-process site consists  in meeting the various constraints coming from measurements of abundances in Galactic stars, from the modeling of possible sources such as supernovae and neutron star mergers, from nuclear physics and, last but not least, from detailed Galactic chemical evolution studies.
   Observations of heavy element abundances in Galactic halo stars provide 
   important constraints on the astrophysical site(s) of r-process 
   nucleosynthesis. Early recognition that the heavy elements abundance 
   patterns in extremely metal-poor stars ([Fe/H]~$< -$3.0 dex) involve only 
   r-process products (Truran 1981) led to the view that r-process 
   nucleosynthesis must be associated with the environments provided by the 
   evolution of massive stars ($m >$~10 M$_\odot$). Neutrino-driven winds from 
   proto-neutron stars following the delayed explosions of very massive stars 
   ($m >$~20 M$_\odot$) have been suggested as a promising site to form the 
   solar r-process abundances, though many shortcomings were immediately 
   apparent with this scenario (Takahashi et al. 1994; Woosley et al. 1994; Wanajo et al. 2001). 
   Proton-rich winds were first found by 
   Liebendoerfer et al. (2003) and 
   discussed in detail also by Froelich et al. (2006a,b) as well as 
   Pruet et al. (2005, 2006).
   Recently, hydrodynamical simulations with accurate neutrino transport have 
   shown that the neutrino winds are proton-rich (Arcones et al. 2007; Fischer 
   et al. 2010; H{\"u}depohl et al. 2010) or slightly neutron-rich 
   (Mart{\'{\i}}nez-Pinedo et al. 2012; Roberts et al. 2012), but never very 
   neutron-rich, as found in older simulations 
(e.g. Woosley et al. 1994; Takahashi et al. 1994). This 
   casts serious doubts on the validity of the neutrino wind scenario. It seems 
   now established that neutrino-driven winds from proto-neutron stars cannot be the 
   main origin of the r-process elements beyond $A \sim$110 (Wanajo 2013; 
   Arcones \& Thielemann 2013). On 
   the other hand, prompt explosions of massive stars in the 8--10~M$_\odot$ 
   range may lead to an ejected amount of r-process matter consistent with the 
   observed Galactic abundances (Wheeler et al. 1998). Yet, it is not clear 
whether 
   these prompt explosions do occur. Recent 3D magneto-hydrodynamics 
   investigations point to supernova progenitors characterized by high rotation 
   rates and large magnetic fields as an interesting site for the strong r-process 
   observed in the early Galaxy (Winteler et al. 2012, and references 
   therein). In this context, the rarity of progenitors with the required 
   initial conditions would also provide a natural explanation for the scatter 
   in the abundances of r-process elements observed for low-metallicity stars. 
   This r-process elements production channel certainly deserves further 
   investigation. 

   Another major source of r-process elements might be neutron star mergers 
   (Lattimer \& Schramm 1974, 1976; Lattimer et al. 1977; Freiburghaus et al. 1999; Goriely et al. 2011; Roberts et al. 2011). The resulting abundance patterns are extremely robust with respect to varying the parameters of the merging binary system  and the results from a double neutron star and a neutron star black hole merger are practically indistinguishable, see, for example, Fig. 4 in Korobkin et al. (2012). Therefore we refer to both types of systems collectively as compact binary mergers (CBM).

   Rosswog et al. (1999, 2000) and more recently Oechslin et al. (2007), Bauswein et al. (2013), Rosswog (2013), Hotokezaka et al. (2013), Kyutoku et al. (2013) showed that up to 10$^{-2}$~M$_\odot$ of r-process matter may 
   be ejected in a single coalescence event. Though this quantity is orders of 
   magnitude higher than the average r-process ejecta required from SNeII, if every 
   SNII is expected to produce r-process matter, the rate of CBM in the Galaxy is 
   significantly lower than the SNII 
   rate, making unclear which one of the sources could potentially be the major 
   r-process elements producer. Neutron star mergers could also provide a natural 
   explanation for the scatter of r-process elements abundances at low metallicity, 
   given their rarity and high r-process element production.

   In principle, chemical evolution studies offer a way to discriminate among 
   different sites for the r-process elements production, through the 
   comparison of model predictions with the observations. Unfortunately, 
   different studies have reached different conclusions (e.g. Travaglio et al. 
   1999; Ishimaru \& Wanajo 1999; De Donder \& Vanbeveren 2003; Argast et al. 
   2004; Cescutti et al. 2006). Furthermore, only rarely all the possible 
   r-process elements sources have been screened in the very same study. In 
   particular, Argast et al. (2004) analysed the possibilities of r-process 
   production from SNeII and CBM, and concluded that it is unlikely that CBM 
   can be responsible for the entire r-process production, because of the delayed Eu 
   appearance predicted by their model, which would not allow to reproduce the Milky 
   Way data at low metallicity. This was due to the adoption of an inhomogeneous   chemical 
   evolution model, dropping the instantaneous mixing approximation (I.M.A.) in the early Galactic 
   evolutionary phases. This model was predicting the spread for r-process elements at low metallicity but also for other elements which do not show a large spread.
   On the other hand, De Donder \& Vanbeveren (2003) had 
   concluded that neutron star (NS)/black hole mergers could be responsible for the Galactic 
   r-process production, but they did not consider the possible contribution 
   from SNeII. Their model assumed I.M.A. 

   In this paper we intend to study again the problem of r-process production 
   in the Galaxy and in particular of europium, in the light of recent and 
   detailed nucleosynthesis calculations of r-process production in CBM 
   (Korobkin et al. 2012), as well as of the existence of new detailed data. We 
   will explore the possibility of Eu production from SNeII and CBM with the 
   intent of establishing which one of the two sources is the most likely and 
   what are the shortcomings of both scenarios. We will adopt a very recent 
   detailed chemical evolution model for the Milky Way including updated 
   stellar yields and reproducing the majority of the observational 
   constraints. We will also predict the expected Eu gradient along the 
   Galactic disc. The paper is organized as follows. In Section~2 we describe 
   the chemical evolution model. In Section~3 we show the adopted Eu 
   nucleosynthesis prescriptions for the CBM and SNeII, as well as the 
   computation of the CBM rate. In Section~4 the results are presented and in 
   Section~5 the main conclusions are summarized.

   \section{The chemical evolution model}
   \label{sec:basic}
 
   We adopted the model~15 of Romano et al. (2010), which contains updated 
   stellar yields and reproduces the majority of the [X/Fe] versus [Fe/H] 
   relations observed in Galactic stars in the solar vicinity. This model is an 
   updated version of the Chiappini et al. (1997, 2001) two-infall model, where 
   it is assumed that the inner halo and part of the thick disc formed by means 
   of a gas accretion episode independently from the thin disc, which formed by 
   means of another gas accretion episode on a much longer timescale. This 
   model relaxes the instantaneous recycling approximation (I.R.A.), i.e. the 
   stellar lifetimes are taken into account in detail, but retains I.M.A., i.e. 
   the stellar ejecta are assumed to cool and mix instantaneously 
   with the surrounding interstellar medium (ISM). 
   The model does not consider Galactic fountains, but see Spitoni et al. (2009), who showed 
   that the fountains should not be affecting the chemical abundances in the disc, as long as 
   the process is shorter than 100 Myr, and does not allow for gas recycling through the hot 
   halo (but see e.g. Brook et al. 2013). In the Spitoni et al. (2009) paper it was shown that the effect of 
   the fountains is also to delay the chemical enrichment, thus breaking the I.M.A.. 
However, it was found that even a delay of several hundreds million years would not change the chemical results for the evolution of the Galactic disc, thus supporting I.M.A.. On the other hand, inhomogeneities in the ISM could be important in the early evolutionary phases, during halo formation. In this respect, there are conflicting results: on one hand it is found a very little spread in abundance ratios of many elements down to [Fe/H]=-4.0 dex, on the other hand  a large spread is observed in abundance ratios involving s- and r-process elements relative to Fe.
In fact, the data for europium show a large spread at low metallicities: the interpretation of this spread, however, can be related more to the different stellar producers of these elements rather than to a less efficient mixing (see  Cescutti et al. 2006, 2013), since in this latter case the spread should be seen for all chemical species. In summary, we think that I.M.A. is not a bad assumption on the scale of the solar vicinity also because, besides the above mentioned results of Spitoni et al. (2009), it has been shown (Recchi et al. 2001) that mixing in the ISM can occur on very short timescales of the order of tens of million years.

   The following equation describes the evolution of the surface mass density 
   of the gas in the form of the generic element $i$, $\sigma_{i}(r,t)$:

   \begin{displaymath}
     \dot{\sigma_{i}}(r,t)=-\psi(r,t)X_{i}(r,t)
   \end{displaymath}
   \smallskip
   \begin{displaymath}
     +\int\limits^{M_{Bm}}_{M_{L}}\psi(r,t-\tau_{m})Q_{mi}(t-\tau_{m})\phi(m)dm
   \end{displaymath}
   \begin{displaymath}
     +A_{Ia}\int\limits^{M_{BM}}_{M_{Bm}}\phi(M_{B}) \cdot
   \end{displaymath} 
   \begin{displaymath}
     \left[\int\limits_{\mu_{m}}^{0.5}f(\mu)\psi(r,t-\tau_{m2})Q^{SNIa}_{mi}(t-\tau_{m2})d\mu\right]dM_{B}
   \end{displaymath}
   \begin{displaymath}
     +(1-A_{Ia})\int\limits^{M_{BM}}_{M_{Bm}}\psi(r,t-\tau_{m})Q_{mi}(t-\tau_{m})\phi(m)dm
   \end{displaymath}
   \begin{displaymath}
     +\int\limits^{M_{U}}_{M_{BM}}\psi(r,t-\tau_{m})Q_{mi}(t-\tau_{m})\phi(m)dm
   \end{displaymath}
   \smallskip
   \begin{equation}
     +X_{A_{i}}A(r,t)
     \label{evol}
   \end{equation}
   where $X_{i}(r,t)$ is the abundance by mass of the element $i$ at the time 
   $t$ and Galactic radius $r$; $Q_{mi}$ indicates the fraction of mass 
   restored by a star of mass $m$ in the form of the element $i$, the so-called 
   `production matrix' as  defined by Talbot \& Arnett (1973). The upper mass 
   limit, $M_{U}$, is set to 100~M$_{\odot}$, while $M_{L}$, the lightest mass 
   which contributes to the chemical enrichment, is set to 0.8~M$_{\odot}$. The 
   parameter $A_{Ia}=0.035$ represents the fraction of binary systems with the 
   right characteristics to give rise to type Ia SNe (SNeIa) in the initial 
   mass function (IMF); its value is chosen in order to obtain the best fit to 
   the present time SNIa rate. The adopted progenitor model for SNeIa is the 
   single-degenerate model as suggested by Greggio \& Renzini (1983) and later 
  reproposed by Matteucci \& Recchi (2001). This formulation, which is different from single-degenerate rates computed by means of population synthesis models (e.g. Mennekens et al. 2010), has proven to be 
   successful in reproducing the chemical evolution of the Milky Way, as well 
   as of other galaxies, such as ellipticals. This rate is also very similar to the rate deriving from the double-degenerate model (see Greggio, 2005) and the two rates produce the same chemical evolution results, as shown in Matteucci et al. (2006, 2009), where the interested reader 
   can find more details. In other words, from the point of view of chemical evolution, using the single- or double-degenerate model or a combination of the two, does not produce any noticeable effect in the chemical results.
   
   The term $A(r,t)$ represents the gas accretion rate:
   \begin{equation}
     A(r,t)= a(r) e^{-t/ \tau_{H}(r)}+ b(r) e^{-(t-t_{max})/ \tau_{D}(r)}.
   \end{equation}
   The quantities $X_{A_{i}}$ are the abundances in the infalling material, 
   which is assumed to be primordial, and are set after Romano et al. (2006), 
   while $t_{max}=1$ Gyr is the time for maximum infall onto the thin disc, 
   $\tau_{H}= 0.8$ Gyr is the time scale for the formation of the inner 
   halo/thick disc and $\tau_{D} (r)$ is the timescale for the formation of the 
   thin disc and is a function of the Galactocentric distance (inside-out 
   formation, Matteucci \& Fran{\c c}ois 1989; Chiappini et al. 2001; 
   Pilkington et al. 2012). In the framework of our model, for the solar 
   neighbourhood the best value is $\tau_{D} (8\,\mathrm{kpc}) =7$~Gyr (see 
   Chiappini et al. 1997; Romano et al. 2010). The quantities $a(r)$ and $b(r)$ 
   are parameters fixed by reproducing the present time total surface mass 
   densities in the halo and disc of the Galaxy (see Romano et al. 2000 for 
   details).

   The adopted IMF, $\phi(m)$, is that of Scalo (1986) and the stellar 
   lifetimes are taken from Schaller et al. (1992). The assumed star formation 
   rate (SFR), $\psi(r,t)$, is a Schmidt-Kennicutt law proportional to the 
   surface gas density to the 1.5th power.

   The model computes in detail the chemical abundances of 37 species in the 
   ISM. For all elements but europium, the adopted stellar yields, that are 
   used to compute the entries of the $Q_{mi}$ matrix, 
   are described in detail in Romano et al. (2010). They reproduce very well 
   the abundance patterns of most chemical species observed in the stars of the 
   Milky Way halo (see also Brusadin et al. 2013) and discs (Micali 
     et al. 2013). 

   In the following, we review a few basic facts.
   \begin{itemize}
     \item Low- and intermediate-mass stars ($m =$ 0.8--8 M$_\odot$) contribute 
       mainly to the chemical enrichment in He, C, N, s-process elements and, 
       perhaps, some $^7$Li and Na. The adopted stellar yields are from Karakas 
       (2010) and rest on detailed stellar evolutionary models.
     \item Massive stars ($m >$~8 M$_\odot$) are responsible for the production 
       of the $\alpha$- and iron-peak elements. The production of (primary) 
       nitrogen and s-process elements is boosted at low-metallicity in fast 
       stellar rotators (Meynet \& Maeder 2002a; Frischknecht, Hirschi \& 
       Thielemann 2012). Here we adopt He, C, N and O yields from pre-supernova 
       models of rotating massive stars from Meynet \& Maeder (2002b), Hirschi 
       et al. (2005), Hirschi (2007) and Ekstr{\"o}m et al. (2008). For heavier 
       elements, yields are from Kobayashi et al. (2006). 
     \item When in binary systems with the right characteristics to give rise 
       to SNIa events, white dwarfs (originating from low- and 
       intermediate-mass stars) are responsible for the production of the bulk 
       of iron in the Galaxy. The adopted SNIa yields are those of Iwamoto et 
       al. (1999, their model W7).
   \end{itemize}
   In the next section, we discuss how europium production from stars has been 
   implemented in our model.

   \section{Europium production sites}

   As discussed in the introduction, two possible sites have been suggested for 
   the production of Eu in stars: SNeII of either low (8--10~M$_{\odot}$) or 
   high mass ($>$~20~M$_{\odot}$; Cowan et al. 1991; Woosley et al. 1994; 
   Wanajo et al. 2001) and CBM (Lattimer \& Schramm 1974; Eichler et al. 1989;  Freiburghaus et al. 1999;
Rosswog et al. 1999, 2000). However, the classical site for the production of r-process elements, 
   namely SNeII, has been recently questioned (e.g. Arcones et al. 2007) while 
   the r-process production in CBM seems a very robust result (e.g. Korobkin 
   et al. 2012).

   We have computed the evolution of the Eu abundance in the Milky Way under 
   several assumptions: i) Eu is produced only in CBM; ii) Eu is produced only 
   in SNII explosions; iii) Eu is produced both in CBM and SNII explosions.

   \subsection{Europium production from compact binary mergers}

   To include the production of Eu from coalescence of neutron stars\footnote{coalescence of a black hole and a NS may work as well}
   Galactic chemical evolution  model we need to define the following 
   quantities:
   \begin{enumerate}
   \item the realization probability for CBM, $\alpha_{\mathrm{CBM}}$;
   \item the time delay between the formation of the double neutron star system 
     and the merging event, $\Delta t_{\mathrm{CBM}}$;
   \item the amount of Eu produced during the merging event, 
     $M^{\mathrm{Eu}}_{\mathrm{CBM}}$.
   \end{enumerate}

   \subsubsection{The neutron star merger rate}

   In our model, the rate of CBM at the time $t$ is computed under the 
   assumption that the rate of formation of double neutron star systems, which 
   will eventually coalesce, is a fraction $\alpha_{\mathrm{CBM}}$ of the 
   neutron star formation rate at the time $t - \Delta t_{\mathrm{CBM}}$:
   \begin{equation}
     R_{\mathrm{CBM}} (t) = \alpha_{\mathrm{CBM}} \cdot \int^{M_{ns,2}}_{M_{ns,1}}
     {\psi(t-\tau_m-\Delta t_{\mathrm{CBM}}) \phi(m) dm},
   \end{equation}
   where $M_{ns,1}=9$  and $M_{ns,2}=30$~M$_{\odot}$ are the canonical lower and upper masses, at birth, which 
   can leave a neutron star as a remnant (we will come back to the issue of the 
   choice of the upper mass limit in Sections~\ref{sec:results} and 
   \ref{sec:discuss}). Stars with $m >$ 30~M$_{\odot}$ probably leave black 
   holes as remnants but the situation is quite uncertain and depends on the 
   assumed rate of mass loss in massive stars and its dependence upon stellar 
   metallicity (e.g. Meynet \& Maeder, 2002a,b). The value of the parameter $\alpha_{\mathrm{CBM}}$ 
   is chosen by imposing that equation~(3) reproduces the present-time rate of 
   neutron star merging in the Galaxy. Several observational estimates of this 
   rate appeared in the literature (van den Heuvel \& Lorimer 1996; Kalogera \& 
   Lorimer 2000; Belczynsky et al. 2002; Kalogera et al. 2004). Here, we take 
   that of Kalogera et al. (2004), $R_{\mathrm{CBM}}(t_{\mathrm{now}}) =$ 
   83$^{+209.1}_{-66.1}$~Myr$^{-1}$, and find $\alpha_{\mathrm{CBM}} =$ 0.018.

   \subsubsection{The time delay}
Based on the energy/angular momentum loss to gravitational waves (Peters and Mathews 1964), inspiral times are usually thought to be between 10 and 100 Myrs, though some studies (e.g. Belczynski et al. 2002) find that a large fraction of systems would merge within less than a 1 Myr.
Argast et al. (2004), in a work similar to 
   ours, considered two different timescales: 1~Myr and 100~Myr. Here we will 
   consider 1~Myr, 10~Myr and 100~Myr. It is worth noting that in both this 
   work and Argast et al. (2004) it is assumed that all neutron star binaries 
   have the same coalescence timescale. Clearly, a more realistic approach 
   would consider a distribution function of such timescales, in analogy with 
   SNeIa for which a distribution for the explosion times is defined (see 
   Greggio 2005). 

   \subsubsection{The Eu yields}

   \begin{table} 
     \caption{Model parameters.}
     \label{tab:par}
     \begin{tabular}{lccc}
       \hline
       Model & $\Delta t_{\mathrm{CBM}}$ & $M^{\mathrm{Eu}}_{\mathrm{CBM}}$ & 
       Yields from type II SNe \\
                     & (Myr) & (M$_\odot$)        & \\
       \hline 
       Mod1NS        & 100 & 10$^{-7}$            & --- \\
       Mod2NS        & 10  & 10$^{-7}$            & --- \\
       Mod3NS        & 1   & 10$^{-7}$            & --- \\
       Mod1NS\arcmin & 100 & 3 $\times$ 10$^{-7}$ & --- \\
       Mod2NS\arcmin & 10  & 3 $\times$ 10$^{-7}$ & --- \\
       Mod3NS\arcmin & 1   & 3 $\times$ 10$^{-7}$ & --- \\
       Mod1NS\arcsec & 100 & 9 $\times$ 10$^{-7}$ & --- \\
       Mod2NS\arcsec & 10  & 9 $\times$ 10$^{-7}$ & --- \\
       Mod3NS\arcsec & 1   & 9 $\times$ 10$^{-7}$ & --- \\
                     &     &                      &     \\
       Mod1SN        & --- & --- &  Cescutti et al. (2006) \\
       Mod2SN        & --- & --- &  Argast et al. (2004)$^a$\\
       Mod3SN        & --- & --- &  Cescutti+Argast$^b$\\
                     &     &                      &      \\  
       Mod1SNNS      & 1   & 10$^{-7}$ & Cescutti et al. (2006) \\
       Mod2SNNS      & 10  & 2 $\times$ 10$^{-7}$ & Argast et al. (2004)$^a$\\
       \hline
     \end{tabular}

     \medskip
     \begin{flushleft}
       $^a$Yields from Table~2 of Argast et al. (2004), their Model~SN2050, but 
       for progenitors in the mass range 20--23~M$_\odot$ a constant yield of 
       3.8~$\times$ 10$^{-8}$ M$_\odot$ is assumed.
       $^b$Yields from Cescutti et al. (2006) in the mass range 12--15~M$_\odot$
       and from Argast et al. (2004; their model SN2050, modified as in Mod2SN) 
       in the mass range 20--50 M$_\odot$.
     \end{flushleft}
   \end{table}

   Every neutron star merging event is assumed to produce the same amount of Eu 
   since we consider only 1.4~M$_\odot$+1.4~M$_\odot$ systems. In the literature 
   there have been different Eu yields reported: Rosswog et al. (1999, 2000) found that up to $10^{-2}$ M$_\odot$  of r-process material are ejected per event and they
pointed out that this would be enough to be a major contribution to the cosmic r-process inventory. A number of recent studies (Oechslin, Janka \& Marek 2007; Rosswog 2013, Bauswein et al. 2013; Hotokezaka et al. 2013; Kyutoku et al. 2013) find a spread of ejecta masses in this range with the exact numbers, around the value of $10^{-2}$, depending on the binary mass ratio, and to some extent on the employed physics.
We take the value of $\sim$~0.01~M$_\odot$ and assume that 
   the mass of Eu is in the range $M_{\mathrm{CBM}}^{\mathrm{Eu}} 
   =$~10$^{-5}$--10$^{-7}$ M$_\odot$, where the lower value is probably the most 
   realistic one. In particular, we compute models (see Table~\ref{tab:par}) 
   assuming different values for the Eu yield: $M_{\mathrm{CBM}}^{\mathrm{Eu}} 
   =$~10$^{-7}$ M$_\odot$, 2~$\times$ 10$^{-7}$ M$_\odot$, 3~$\times$ 10$^{-7}$ 
   M$_\odot$ and 9~$\times$ 10$^{-7}$ M$_\odot$.

   \subsection{Europium production from core-collapse supernovae}
   \label{sec:euii}

   The yields of r-process elements and therefore of Eu from SNeII are highly 
   uncertain. Cescutti et al. (2006) suggested some empirical Eu yields 
   dictated by the need of reproducing the trend of [Eu/Fe] versus [Fe/H] 
   observed for Galactic stars. They suggested that Eu is a pure r-process 
   element and that it is produced in the mass range 12--30~M$_\odot$. In 
   particular, according to Cescutti et al. (2006), a 12~M$_\odot$ star must 
   produce $M^{\mathrm{Eu}}_{12} =$ 4.5~$\times$ 10$^{-8}$~M$_\odot$, a 
   15~M$_\odot$ star, $M^{\mathrm{Eu}}_{15} =$ 3.0~$\times$ 10$^{-8}$~M$_\odot$ 
   and a 30~M$_\odot$ star, $M^{\mathrm{Eu}}_{30} =$ 5.0~$\times$ 
   10$^{-10}$~M$_\odot$. Argast et al. (2004) adopted somewhat different 
   empirical yields: they considered either the lower-mass SNeII 
   (8--10~M$_\odot$) or the higher-mass SNeII (20--50~M$_\odot$) as dominant 
   r-process sites. In this paper, we show the results of our model adopting 
   either the yields from Cescutti et al. (2006) or the yields from Argast et 
   al. (2004; see Table~\ref{tab:par}). In particular, from the latter we adopt 
   the yields corresponding to their Model~SN2050 (see their Table~2), but for 
   progenitors in the mass range 20--23~M$_\odot$ we assume a constant yield of 
   3.8~$\times$ 10$^{-8}$ M$_\odot$ rather than a declining yield from 
   1.8~$\times$ 10$^{-6}$ M$_\odot$ to 3.8~$\times$ 10$^{-8}$ M$_\odot$, as in 
   the original paper. This assumption is required in order to fit the data, as 
   we will see in the next section.

   \section{Model results}
   \label{sec:results}

   We run several models with neutron stars as the only Eu source, models with 
   SNeII as the only Eu source as well as  models with both sources acting at 
   the same time. We vary the yield from merging neutron stars as well as the 
   time delay for merging and the yields from SNeII. In Table~\ref{tab:par} we 
   show the model parameters: in column~1 we report the name of the model, in 
   column~2 the assumed coalescence timescale, in column~3 the assumed yield 
   for CBM and in column~4 the literature source for the assumed yields for 
   SNeII. All models assume the same SFR and the same IMF (see 
   Sect.~\ref{sec:basic}).


   \begin{figure}
   \psfig{figure=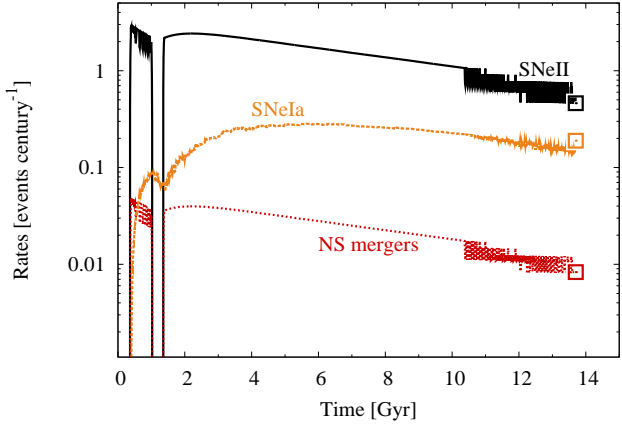,width=\columnwidth}
     \caption{Predicted SNII, SNIa and CBM rates (black solid, orange 
       dashed and red dotted curves, respectively) as functions of cosmic time for the Milky Way. 
       Also shown are the observational present-time values (squares; SN rates: 
       Li et al. 2011; CBM rate: Kalogera et al. 2004).}
     \label{fig:rates}
   \end{figure}


   In Fig.~\ref{fig:rates} we show the predicted SNII and SNIa rates, which are 
   common to all the models, and the predicted CBM rate. As one can see, the 
   CBM rate follows strictly the SNII rate in shape, although it is smaller in 
   absolute value. In the same figure we report the observed values of the 
   three rates at the present time in the Galaxy and the agreement between 
   model predictions and observations is good. 


   \begin{figure} 
     \psfig{figure=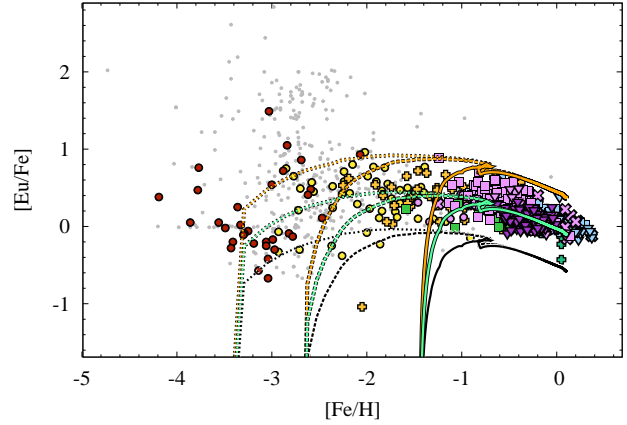,width=\columnwidth}
     \caption{Predicted and observed [Eu/Fe] versus [Fe/H] relations for solar 
       neighbourhood stars. The dotted lines all refer to models with $\Delta 
       t_{\mathrm{CBM}}=$ 1~Myr, the dashed ones to models with $\Delta 
       t_{\mathrm{CBM}}=$ 10~Myr and the solid ones to models with $\Delta 
       t_{\mathrm{CBM}}=$ 100~Myr. The lowermost (black) lines refer to models 
       that assume $M^{\mathrm{Eu}}_{\mathrm{CBM}}=$ 10$^{-7}$~M$_\odot$, the 
       middle (green) lines to models that assume 
       $M^{\mathrm{Eu}}_{\mathrm{CBM}}=$ 3~$\times$ 10$^{-7}$~M$_\odot$ and the 
       uppermost (orange) ones to models that assume 
       $M^{\mathrm{Eu}}_{\mathrm{CBM}}=$ 9~$\times$ 10$^{-7}$~M$_\odot$ (see 
       Table~\ref{tab:par}). Data (circles are for halo stars, squares for 
       thick-disc members, upside-down triangles for thin-disc members, crosses 
       for transition objects and `+' signs for objects with no assigned 
       kinematic membership) are from: Burris et al. (2000; yellow circles); 
       Fulbright (2000; orange `+' signs); Reddy et al. (2003; purple 
       upside-down triangles); Bensby et al. (2005; light blue symbols); Reddy 
       et al. (2006; magenta symbols); Fran\c cois et al. (2007; red circles, mainly upper limits); 
       Mishenina et al. (2007; green `+' signs); Ramya et al. (2012; green 
       squares). The small (grey) dots are data from the compilation of 
       metal-poor stars of Frebel (2010).}
     \label{fig:NS}
   \end{figure} 


   In Fig.~\ref{fig:NS} we show the predicted (lines) and observed (symbols) 
   [Eu/Fe] versus [Fe/H] relations: the model predictions refer to the three 
   sets of models with Eu production only from CBM (see Table~\ref{tab:par} 
   from Mod1NS to Mod3NS\arcsec).  The data are a large compilation including 
   the most recent ones. It is clear from this figure that if 
   $\Delta t_{\mathrm{CBM}} \ge$ 10~Myr the CBM cannot explain the [Eu/Fe] 
   ratios at low [Fe/H]. In any case, even if the minimum value for 
   $\Delta t_{\mathrm{CBM}}$ is assumed, i.e. 1~Myr, it is not possible to 
   explain the observed points at [Fe/H]$< -$3.5 dex. This suggests that 
   another Eu source should be active on very short timescales, such as SNeII 
   of high mass (up to 50~M$_\odot$). Moreover, if the CBM are the only source 
   of Eu, then the best yield should be 3~$\times$ 10$^{-7}$~M$_\odot$, which 
   is the value that best fits the average trend of the data in 
   Fig.~\ref{fig:NS}. In fact, although a spread is present in the data, 
   especially at low metallicities, our model is aimed at fitting the average 
   trend. 


   \begin{figure} 
     \psfig{figure=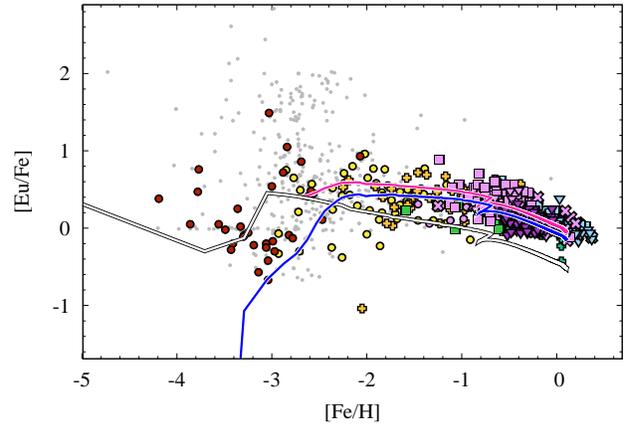,width=\columnwidth}
     \caption{Predicted and observed [Eu/Fe] versus [Fe/H] relations for solar 
       neighbourhood stars. The blue, white and magenta curves refer to the 
       predictions of models Mod1SN, Mod2SN and Mod3SN, respectively (see 
       text). Data as in Fig.\ref{fig:NS}.}
     \label{fig:SN}
   \end{figure} 


   In Fig.~\ref{fig:SN} we show the results of the three models with Eu production 
   only from SNe~II. We consider two different yield sets (see 
   Table~\ref{tab:par}, models labelled Mod1SN, Mod2SN and Mod3SN). The model 
   with the yields from Cescutti et al. (2006; model Mod1SN) fits reasonably 
   well the data for [Fe/H]~$\ge -$2.0 dex, but it does not explain the [Eu/Fe] 
   ratios in stars at lower metallicities. The model with the yields from 
   Argast et al. (2004; model Mod2SN) is able to reproduce the low-metallicity 
   data, but fails to reproduce the observations for [Fe/H]~$> -$2.0 dex. 
   Moreover, in order not to overproduce Eu at [Fe/H]~$\le -$3.0 dex, we have 
   to reduce the original yields in the 20--23~M$_\odot$ mass range (see 
   before). In fact, the kink at [Fe/H]$\sim$ -3.0 dex in the white curve
is related to the appearance of SNeII with progenitors with initial mass $\sim 20$ M$_{\odot}$.
We had to reduce the Eu produced by a  $\sim 20$ M$_{\odot}$ from $10^{-5}$ M$_{\odot}$ to
$3.8 \cdot 10^{-8}$ M$_{\odot}$ in order to best fit the data and reduce the kink.

\begin{figure}
     \psfig{figure=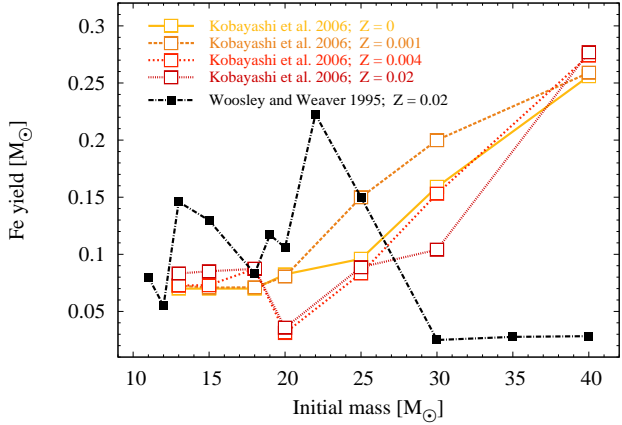,width=\columnwidth}
     \caption{ Iron yields from core-collapse SNe, as a function of the initial 
       mass of the progenitor, according to different nucleosynthesis studies, 
       and for different values of the initial metallicities of the stars.}
     \label{fig:feyields}
   \end{figure}

   From the comparison of our model predictions with the observations, 
   we deduce that the yields from SNeII should be adjusted and that the mass 
   range of Eu producers should be extended up to 50~M$_\odot$. This is 
   dictated by the most recent data on [Eu/Fe] extending down to 
   [Fe/H]~$\le-$4.0~dex. Cescutti et al. (2006) had derived the range 
   12--30~M$_\odot$ for Eu producers, by comparison with older data and by 
   adopting different Fe yields from massive stars than in this paper. In 
   particular, Cescutti et al. (2006) adopted the Fe yields of Woosley \& 
   Weaver (1995) for solar metallicity, whereas here we adopt the more recent 
   yields depending on metallicity by Kobayashi et al. (2006). These latter 
   predict a higher increase of the Fe abundance in the early Galactic 
   phases relative to the Woosley \& Weaver (1995) yields. A comparison of 
     the Fe yields adopted in Cescutti et al. (2006) and our study is shown in 
     Fig.~\ref{fig:feyields}. We also deduce that stars with masses smaller 
   than 20~M$_\odot$ must contribute to Eu production, in order to 
   counterbalance the production of Fe from SNeIa (see Fig.~\ref{fig:SN}, white 
   solid line). Therefore, we run a hybrid model (model Mod3SN; see 
   Table~\ref{tab:par} and Fig.~\ref{fig:SN}, magenta line) that uses the 
   yields from Cescutti et al. (2006), but only for progenitors in the 
   12--15~M$_\odot$ mass range, and those from Argast et al. (2004) for 
   progenitors in the 20--50~M$_\odot$ mass range (we reduce the original 
   yields in the 20--23~M$_\odot$ mass range ---see Notes to 
   Table~\ref{tab:par}). This model reproduces very well the data over the full 
   metallicity range.


   \begin{figure}
     \psfig{figure=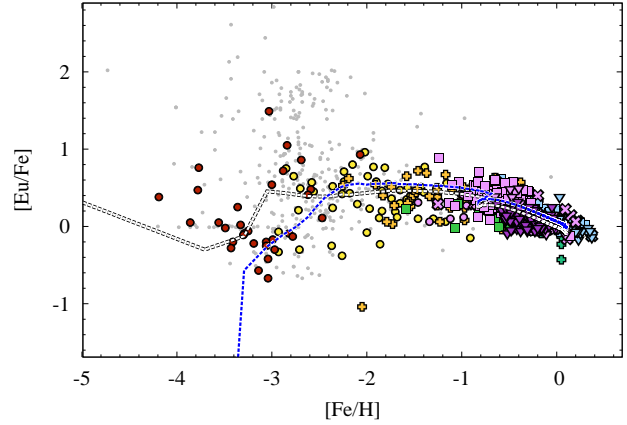,width=\columnwidth}
     \caption{Same as Figs.\ref{fig:NS} and \ref{fig:SN}, but now the 
       theoretical predictions refer to models Mod1SNNS and Mod2SNNS (blue and 
       white curves, respectively), that account for both CBM and SNeII 
       as Eu factories (see text).}
     \label{fig:SNNS}
   \end{figure} 


   In Fig.~\ref{fig:SNNS} we show the results of the models with contributions 
   to Eu synthesis from both SNeII and CBM. Model Mod1SNNS assumes the SNII 
   yields from Cescutti et al. (2006) and the lowest Eu production from CBM 
   (10$^{-7}$~M$_{\odot}$) with a time delay of 1~Myr (see 
   Table~\ref{tab:par}). 
   Also in this case, like for the models presented in Fig.~\ref{fig:NS}, we 
   can say that the minimum coalescence timescale (1~Myr) is required to fit 
   the data, but still unsuited to reproduce the data for very low-metallicity 
   stars. Model Mod2SNNS assumes modified Eu yields for supernovae from Argast et al. (2004); 
   the adopted Eu yield from CBM is slightly higher than in model Mod1SNNS 
   (2~$\times$ 10$^{-7}$~M$_\odot$), while the coalescence timescale is longer 
   (10~Myr). It is shown that the joint contribution to Eu synthesis from both 
   high-mass SNeII and CBM (whose progenitors are in the range 
   9--30~M$_\odot$) guarantees a good fit to the available data across the full 
   metallicity range.

   \begin{table} 
     \caption{Solar abundances.}
     \label{tab:sun}
     \begin{tabular}{lcc}
       \hline
       Model & $X_{\mathrm{Fe}, \odot}$ & $X_{\mathrm{Eu}, \odot}$ \\
       \hline 
       Mod1NS           & 1.35~$\times$ 10$^{-3}$ & 1.16~$\times$ 10$^{-10}$ \\
       Mod2NS           & 1.35~$\times$ 10$^{-3}$ & 1.15~$\times$ 10$^{-10}$ \\
       Mod3NS           & 1.35~$\times$ 10$^{-3}$ & 1.15~$\times$ 10$^{-10}$ \\
       Mod1NS\arcmin    & 1.35~$\times$ 10$^{-3}$ & 3.48~$\times$ 10$^{-10}$ \\
       Mod2NS\arcmin    & 1.35~$\times$ 10$^{-3}$ & 3.46~$\times$ 10$^{-10}$ \\
       Mod3NS\arcmin    & 1.35~$\times$ 10$^{-3}$ & 3.46~$\times$ 10$^{-10}$ \\
       Mod1NS\arcsec    & 1.35~$\times$ 10$^{-3}$ & 1.05~$\times$ 10$^{-9}$  \\
       Mod2NS\arcsec    & 1.35~$\times$ 10$^{-3}$ & 1.04~$\times$ 10$^{-9}$  \\
       Mod3NS\arcsec    & 1.35~$\times$ 10$^{-3}$ & 1.04~$\times$ 10$^{-9}$  \\
                        &                         &                         \\
       Mod1SN           & 1.35~$\times$ 10$^{-3}$ & 3.19~$\times$ 10$^{-10}$ \\
       Mod2SN           & 1.35~$\times$ 10$^{-3}$ & 1.35~$\times$ 10$^{-10}$ \\
       Mod3SN           & 1.35~$\times$ 10$^{-3}$ & 4.02~$\times$ 10$^{-10}$ \\
                        &                         &                         \\
       Mod1SNNS         & 1.35~$\times$ 10$^{-3}$ & 4.34~$\times$ 10$^{-10}$ \\
       Mod2SNNS         & 1.35~$\times$ 10$^{-3}$ & 3.65~$\times$ 10$^{-10}$ \\
       \hline
     \end{tabular}
   \end{table}

   In Table~\ref{tab:sun} we list the predicted Eu and Fe solar abundances by 
   mass from the various models. They correspond to the abundances in the ISM 
   4.5 Gyr ago. The observed values for the Eu and Fe solar abundances are 
   $X_{\mathrm{Eu}} =$ 3.5~$\times$ 10$^{-10}$ and $X_{\mathrm{Fe}} =$ 
   1.34~$\times$ 10$^{-3}$, respectively (Asplund et al. 2009). Most of the 
   models predict Eu solar abundances in agreement with the observed one.

   \subsection{The mass range for NS progenitors}

\begin{figure}
     \psfig{figure=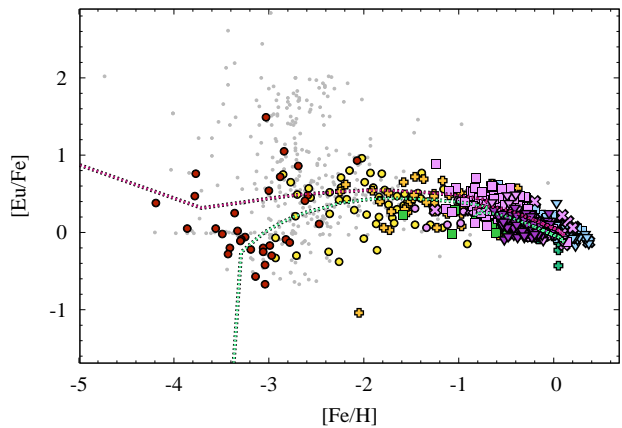,width=\columnwidth}
     \caption{Predicted (curves) and observed (symbols) [Eu/Fe] versus [Fe/H] 
       relations for solar neighbourhood stars. Model predictions refer to model~Mod3NS\arcmin assuming either 30 (green dotted line) or 50~M$_\odot$ (purple dotted line) as the limiting mass for neutron star formation.}
     \label{fig:NS50}
   \end{figure}

   Since in all the previous models we fixed the maximum mass giving rise to a 
   neutron star to be 30~M$_\odot$ and this value is quite uncertain, we 
   decided to try to change this upper limit. In Fig.\ref{fig:NS50} we show the 
   effect on model predictions of a different choice of the upper mass limit 
   for neutron star formation in equation~(3), namely, 50~M$_\odot$, rather 
   than 30~M$_\odot$. It is shown that in this case CBM alone can, in 
   principle, account for the abundances of Eu observed in Galactic halo 
   stars, as well as for the solar Eu (see Table 2).



   \begin{figure}
     \psfig{figure=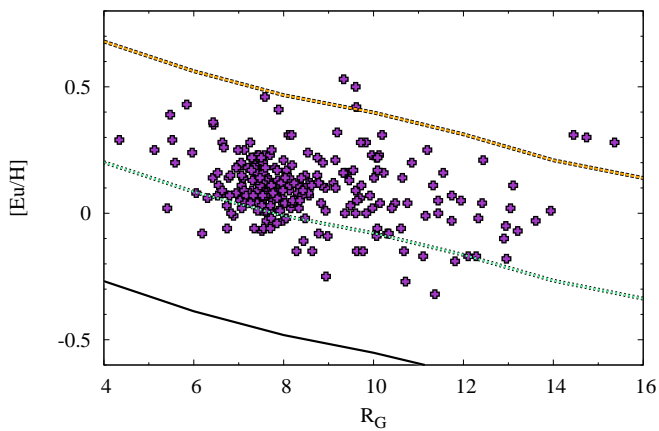,width=\columnwidth}
     \caption{Current radial behaviour of [Eu/H] predicted by models Mod1NS 
     (solid black line), Mod2NS\arcsec (orange dashed line) and Mod3NS\arcmin 
     (green dotted line). Data (crosses) are Cepheids from Luck et al. (2011).}
     \label{fig:grad}
   \end{figure} 



   \begin{figure}
     \psfig{figure=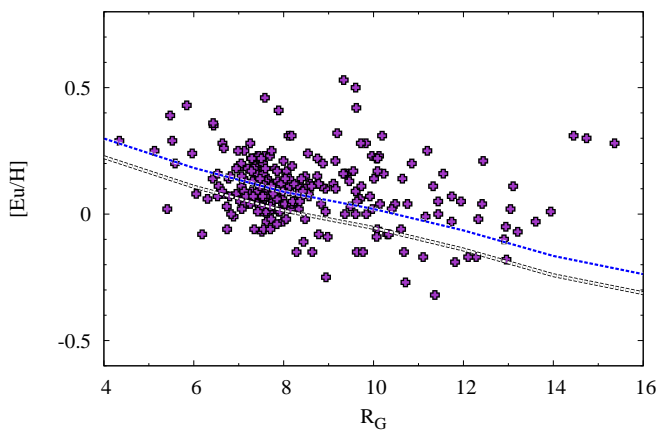,width=\columnwidth}
     \caption{Same as Fig.~\ref{fig:grad}, but here we show the predictions 
       from models Mod1SNNS (blue dashed curve) and Mod2SNNS (white dashed 
       curve).}
     \label{fig:gradB}
   \end{figure} 


\subsection{Eu gradient along the Galactic disc}

   Finally, in Figs.~\ref{fig:grad} and \ref{fig:gradB} we show the predicted 
   gradient of Eu along the Galactic disc at the present time for a subset of 
   models. In both figures, the data are from Cepheids by Luck et al. (2011). 
   In Fig.~\ref{fig:grad} we show cases with Eu production only by CBM. In 
   Fig.~\ref{fig:gradB} we show cases with Eu production from both neutron 
   stars and SNeII. Both scenarios can account reasonably well for the observed 
   gradient of europium in the Milky Way disc. This is because the gradient 
   depends on the mechanism of formation of the disc, here assumed to be the 
   inside-out one. This mechanism was suggested by Matteucci \& Fran\c cois 
   (1989) and proven to be valid also in semi-analytical models (e.g. 
   Pilkington et al. 2012). What it changes in Figs~\ref{fig:grad} and 
   \ref{fig:gradB} is the absolute value of the Eu abundance, [Eu/H], as due 
   to the different assumptions made on the Eu producers. In this case, the 
   models with both SNeII and CBM as Eu producers produce the best agreement 
   with the data (see Fig.~\ref{fig:gradB}).

   \section{Conclusions and discussion}
   \label{sec:discuss}

   In this paper we have studied the production of Eu and assumed that this 
   element can be produced in both compact binary mergers (CBM) and SNeII. To do that, we have adopted a 
   very detailed chemical evolution model which can predict the evolution of 
   the abundances of many species and that already reproduces the behaviour of 
   several abundances as well as the main features of the solar neighbourhood 
   and the whole disc. Our attention has been focused here on the production of 
   Eu by CBM and whether this production alone can explain the solar abundance 
   of Eu as well as the  [Eu/Fe] versus [Fe/H] relation. The main parameters 
   involved in Eu production from CBM are: i) the Eu yield, ii) the time 
   required for the binary neutron star system to coalesce and iii) the range 
   of progenitors of neutron stars. Among these parameters, ii) and iii) are 
   quite uncertain whereas the yields seem to be more reliable.

   Our main conclusions can be summarized as follows:
   \begin{itemize}
   \item CBM can be entirely responsible for Eu production in the Galaxy if 
     the NS systems all have a coalescence time scale no longer than 1 Myr, a 
     possibility suggested by Belczynski et al. (2002), each event produces at 
     least 3~$\times$ 10$^{-7}$~M$_{\odot}$ of Eu and all stars with masses in 
     the range 9--50~M$_{\odot}$ leave a NS as a remnant. In this case we can 
     well reproduce the average trend of [Eu/Fe] versus [Fe/H] in the solar 
     vicinity as well as the Eu solar abundance. In this case, there is no need 
     for SNeII producing Eu. However, all the uncertainties in these parameters 
     plus the uncertain observed rate of CBM in the Galaxy at the present time, 
     induce some caution in drawing firm conclusions.
   \item Perhaps, a more realistic situation would be the one where both CBM 
     and SNeII are producing Eu. The best model in this case requires that CBM 
     can produce 2~$\times$ 10$^{-7}$~M$_{\odot}$ of Eu and the delay times can 
     be various, spanning between 10 and 100 Myr. SNeII should then produce Eu 
     in the range 20--50~M$_{\odot}$ with yields of the order of 
     10$^{-8}$--10$^{-9}$~M$_{\odot}$ of Eu per supernova. It is very important 
     to have high stellar masses to produce the Eu observed at very low 
     metallicity.
   \item Both models with Eu produced only by CBM and models with CBM and SNe 
     can reproduce the Eu abundance gradient observed along the Galactic thin 
     disc.
   \end{itemize}

   Our conclusions are different from those of Argast et al. (2004) who 
   concluded that CBM cannot be the only Eu producers. We think that this is 
   due to the fact that Argast et al.'s model does not assume instantaneous 
   mixing in the early Galactic evolutionary phases. This fact leads to an 
   additional delay in the appearence of Eu in the ISM, besides the delay for 
   the merging of the two NS, and is probably the reason why their predicted 
   [Eu/Fe] appears at a too high [Fe/H] values relative to observations. On the 
   other hand, the assumption that the ISM was not well mixed at early times 
   can explain the large spread observed in the r- and s-process abundances 
   relative to Fe at low metallicity. However, such a large spread is not 
   observed for other abundance ratios at the same metallicity. Therefore, 
   either the spread is explained as due to different stellar producers of 
   different elements as suggested in Cescutti et al. (2006, 2013) or the 
   spread is related to observational errors. Inhomogeneous mixing, in fact, 
   should act on all the elements. Our model assumes instantaneous mixing approximation (I.M.A.) and therefore it 
   cannot reproduce the observed spread but just the average trends.

   On the other hand, De Donder \& Vanbeveren (2003), using a model similar to 
   this one with I.M.A., taken from Chiappini et al. (1997), and computing 
   population synthesis binary models, explored several cases of mergers: 
   NS/NS and NS/black hole. They concluded that mergers NS/black hole can 
   produce enough Eu by themselves but they did not test the case of Eu 
   production from SNeII. More recently, Mennekens \& Vanbeveren (2013), 
   adopting again a population synthesis model and a Galactic model like in De 
   Donder \& Vanbeveren (2003), reached a conclusion similar to that of the 
   present paper: the CBM can account for the entire r-process production 
   except in the first 100 Myr, but they did not include the contribution from 
   SNeII.
   Therefore, although firm conclusions on the nature of Eu cannot be yet drawn,
   this subject should be pursued by testing the various hypothesis discussed here in a model which takes into account early inhomogeneities
as well as a distribution function of the delay times for the merging of neutron stars, and that will be 
the subject of a
forthcoming paper. 

   \section*{Acknowledgements}

   F.M. and D.R. acknowledge financial support from PRIN MIUR~2010--2011, project 
   ``The Chemical and Dynamical Evolution of the Milky Way and Local Group 
   Galaxies'', prot.~2010LY5N2T. 
A.A. aknowledges financial support from the Helmoltz-University Young Investigator grant no. VH-NG-825. 
O.K. and S.R. have been supported by  by DFG grant RO-3399, AOBJ-584282  and by the Swedish Research Council (VR) under grant 621- 2012-4870. S.R. has also been supported by Compstar.
Finally, we thank a highly competent referee for his/her careful reading of the manuscript and useful suggestions.

   \bsp

   \label{lastpage}


\begin{thebibliography}{90}
\bibitem{}
Arcones A., Janka H.-Th., Scheck L., 2007, A\&A, 467, 1227
\bibitem{} Arcones, A., Thielemann, F.-K., 2013, JPhG, 40, 3201
\bibitem{}
Argast D., Samland M., Thielemann F.-K., Qian Y.-Z., 2004, A\&A, 416, 997
\bibitem{} Asplund, M., Grevesse, N., Sauval, A. J., Scott, P., 2009, ARA\&A, 47, 481
\bibitem{}
Bauswein, A., Goriely, S., Janka, H.-T., 2013, ApJ, 773, 78
\bibitem{} Bensby, T., Feltzing, S., Lundstr\"om, I., Ilyin, I., 2005, A\&A, 433,185
\bibitem{} Belczynski, K., Kalogera, V., Bulik, T., 2002, ApJ, 572, 407
\bibitem{}
Brook C. B., Stinson G., Gibson B. K., Shen S., Macci{\`o} A. V., Wadsley J., 
Quinn T., 2013, preprint (arXiv:1306.5766)
\bibitem{}
Brusadin G., Matteucci F., Romano D., 2013, A\&A, 554, A135
\bibitem{}
Burbidge E.~M., Burbidge G.~R., Fowler W.~A., Hoyle F., 1957, Rev. Modern 
Phys., 29, 547
\bibitem{} Burris, D.L., Pilachowski, C. A., Armandroff, T. E., Sneden, C., Cowan, J. J., Roe, H., 2000,
ApJ, 544, 302
\bibitem{}
Cescutti G., Fran{\c c}ois P., Matteucci F., Cayrel R., Spite M., 2006, A\&A, 
448, 557
\bibitem{}
Cescutti, G., Chiappini, C., Hirschi, R., Meynet, G., Frischknecht, U. 2013, A\&A, 553, 51
\bibitem{}
Chiappini C., Matteucci F., Gratton R., 1997, ApJ, 477, 765
\bibitem{}
Chiappini C., Matteucci F., Romano D., 2001, ApJ, 554, 1044
\bibitem{}
Cowan J. J., Thielemann F.-K., Truran J. W., 1991, Phys. Rep., 208, 267
\bibitem{} De Donder, E. \& Vanbeveren, D., 2003, NewAstr., 8, 415

\bibitem{} Eichler D., Livio M., Piran T., Schramm D.N., 1989, Nature,340, 126
\bibitem{}Ekstr{\"o}m S., Meynet G., Chiappini C., Hirschi R., Maeder A., 2008, A\&A, 
489, 685
\bibitem{}
Fischer T., Whitehouse S. C., Mezzacappa A., Thielemann F.-K., Liebend{\"o}rfer 
M., 2010, A\&A, 517, A80

\bibitem{} François, P., Depagne, E., Hill, V., Spite, M., Spite, F., Plez, B., Beers, T. C., Andersen, J.
et al.,  2007, A\&A, 476, 935
\bibitem{} Frebel, A.,  2010, Astronomische Nachrichten, Vol.331, Issue 5, p.474-488 
\bibitem{}
Freiburghaus C., Rosswog S., Thielemann F.-K., 1999, ApJ, 525, L121
\bibitem{}
Frischknecht U., Hirschi R., Thielemann F.-K., 2012, A\&A, 538, L2
\bibitem{} Froehlich, C.,  Hix, W. R., Martínez-Pinedo, G., Liebendoerfer, M., Thielemann, F.-K., Bravo, E., Langanke, K., Zinner, N. T., 2006a, NewAR, 50, 496
\bibitem{} Froehlich, C., Martínez-Pinedo, G., Liebendoerfer, M., Thielemann, F.-K., Bravo, E., Hix, W. R., Langanke, K., Zinner, N. T., 2006b, PhRvL, 96, 2502
\bibitem{} Fulbright, J.P.,  2000, AJ, 120, 1841
\bibitem{} Goriely, S., Bauswein, A., Janka, H.-T., 2011, ApJ, 738, L32
\bibitem{} Greggio, L., 2005, A\&A, 441, 1055
\bibitem{}
Greggio L., Renzini A., 1983, A\&A, 118, 217
\bibitem{}
Hirschi R., 2007, A\&A, 461, 571
\bibitem{}
Hirschi R., Meynet G., Maeder A., 2005, A\&A, 433, 1013
\bibitem{} Hotokezaka, K. et al., 2013, PhRvD, 88, 4026
\bibitem{}
H{\"u}depohl L., M{\"u}ller B., Janka H.-T., Marek A., Raffelt G. G., 2010, 
Phys. Rev. Lett., 104, 251101
\bibitem{} Ishimaru, Y. \& Wanajo, S., 1999, ApJ, 511, L33
\bibitem{}
Iwamoto K., Brachwitz F., Nomoto K., Kishimoto N., Umeda H., Hix W. R., 
Thielemann F.-K., 1999, ApJS, 125, 439
\bibitem{} Kalogera V., Kim C., Lorimer D. R., Burgay M., D'Amico
N., Possenti A., Manchester R. N., Lyne A. G., Joshi
B. C., McLaughlin M. A., Kramer M., Sarkissian J. M.,
Camilo F., 2004, ApJ, 614, L137

\bibitem{} Kalogera, V. \& Lorimer, D.R:  2000 ApJ, 530, 890
\bibitem{} Karakas, A.I.,  2010, MNRAS, 403, 1413
\bibitem{}
Kobayashi C., Umeda H., Nomoto K., Tominaga N., Ohkubo T., 2006, ApJ, 653, 1145
\bibitem{} Korobkin O., Rosswog S., Arcones A., Winteler C., 2012, MNRAS, 426, 1940
\bibitem{} Kyutoku, K., Ioka, K., Shibata, M., 2013, PhRvD, 88, 1503

\bibitem{} Li, W., et al. 2011, MNRAS, 412, 1473
\bibitem{} Luck, R. E., Andrievsky, S. M., Kovtyukh, V. V., Gieren, W.; Graczyk, D. 2011, AJ, 142, 51
\bibitem{} Lattimer J. M., Mackie F., Ravenhall D. G., Schramm D. N., 1977, ApJ, 213, 225
\bibitem{}Lattimer, J. M.\& Schramm, D. N., 1974, ApJ, 192, L145
\bibitem{}Lattimer, J. M.\& Schramm, D. N., 1976, ApJ, 210, 549
\bibitem{}  Liebendoerfer,  M., Mezzacappa, A., Messer, O. E. B., Martinez-Pinedo, G., Hix, W. R., Thielemann, F.-K., 2003, NuPhA, 719, 144

\bibitem{}
Mart{\'{\i}}nez-Pinedo G., Fischer T., Lohs A., Huther L., 2012, Phys. Rev. 
Lett., 109, 251104
\bibitem{}
Matteucci F., Fran{\c c}ois P., 1989, MNRAS, 239, 885
\bibitem{}
Matteucci F., Recchi S., 2001, ApJ, 558, 351
\bibitem{}
Matteucci F., Panagia N., Pipino A., Mannucci F., Recchi S., Della Valle M., 
2006, MNRAS, 372, 265
\bibitem{}
Matteucci F., Spitoni E., Recchi S., Valiante R., 2009, A\&A, 501, 531
\bibitem{} Mennekens N., Vanbeveren, D., De Greve, J.P., De Donder, E., 2010, A\&A, 515, 89
\bibitem{} Mennekens, N. \& Vanbeveren 2013 astro-ph/1307.0959, A\&A submitted
\bibitem{}
Meynet G., Maeder A., 2002a, A\&A, 381, L25
\bibitem{}
Meynet G., Maeder A., 2002b, A\&A, 390, 561
\bibitem{}
Micali A., Matteucci F., Romano D., 2013, preprint (arXiv:1309.1283), MNRAS in press
\bibitem{} Mishenina, T. V., Gorbaneva, T. I., Bienaymé, O., Soubiran, C., Kovtyukh, V. V., Orlova, L. F., 2007, Astronomy Reports, Volume 51, Issue 5, p.382
\bibitem{}Oechslin, R., Janka, H.-T., Marek, A., 2007, A\&A, 467, 395

\bibitem{} Peters, P. \& Mathews, J., 1964, Phys. Review B, 27, 173001
\bibitem{}
Pilkington K., et al., 2012, A\&A, 540, A56
\bibitem{}
Portinari L., Chiosi C., Bressan A., 1998, A\&A, 334, 505
\bibitem{} Pruet, J., Woosley, S. E., Buras, R., Janka, H.-T., Hoffman, R. D., 2005, ApJ, 623, 325
\bibitem{} Pruet, J., Hoffman, R. D., Woosley, S. E., Janka, H.-T., Buras, R., 2006, ApJ, 644, 1028
\bibitem{} Ramya, P., Reddy, B. E., Lambert, D. L. 2012, MNRAS, 425, 3188
\bibitem{} Reddy, B. E., Tomkin, J., Lambert, D. L., Allende Prieto, C.,  2003, MNRAS, 340, 304
\bibitem{} Recchi, S., Matteucci, F., D'Ercole, A., 2001, MNRAS, 322, 800
\bibitem{} Reddy B. E., Lambert D. L., Allende Prieto C., 2006, MNRAS, 367, 1329

\bibitem{} Roberts, L. F., Kasen, D., Lee, W. H., Ramirez-Ruiz, E., 2011, ApJ, 736, L21
\bibitem{}
Roberts L. F., Reddy S., Shen G., 2012, Phys. Rev. C, 86, 065803
\bibitem{}
Romano D., Karakas A. I., Tosi M., Matteucci F., 2010, A\&A, 522, A32
\bibitem{}
Romano D., Matteucci F., Salucci P., Chiappini C., 2000, ApJ, 539, 235
\bibitem{}
Romano D., Tosi M., Chiappini C., Matteucci F., 2006, MNRAS, 369, 295
\bibitem{} Rosswog, S., 2013, Philosophical Transactions of the Royal Society A: Mathematical, Physical and Engineering Sciences, vol. 371, issue 1992, pp. 20120272-20120272

\bibitem{}
Rosswog S., Davies M. B., Thielemann F.-K., Piran T., 2000, A\&A, 360, 171
\bibitem{}
Rosswog S., Liebend{\"o}rfer M., Thielemann F.-K., Davies M. B., Benz W., 
Piran T., 1999, A\&A, 341, 499
\bibitem{}
Scalo J. M., 1986, Fundam. Cosm. Phys., 11, 1
\bibitem{}
Schaller G., Schaerer D., Meynet G., Maeder A., 1992, A\&AS, 96, 269
\bibitem{}
Seeger P. A., Fowler W. A., Clayton D. D., 1965, ApJS, 11, 121
\bibitem{}
Spitoni E., Matteucci F., Recchi S., Cescutti G., Pipino A., 2009, A\&A, 504, 
87
\bibitem{}Takahashi, K., Witti, J., Janka, H.-T., 1994, A\&A, 286, 857

\bibitem{}
Talbot R. J. Jr., Arnett W. D., 1973, ApJ, 186, 51
\bibitem{}
Thielemann F.-K., et al., 2010, J. Phys.: Conf. Ser., 202, 012006
\bibitem{}
Thielemann F.-K., K\"appeli R., Winteler C., Perego A., Liebend{\"o}rfer M., 
Nishimura N., Vasset N., Arcones A., 2012, Proceedings of the XII International 
Symposium on Nuclei in the Cosmos, published online at at http://pos.sissa.it/cgi-bin/reader/conf.cgi?confid=146, id.61
\bibitem{} Travaglio, C., Galli, D., Gallino, R., Busso, M., Ferrini, F., Straniero, O., 1999, ApJ, 521, 691
\bibitem{}
Truran J. W., 1981, A\&A, 97, 391
\bibitem{} van den Heuvel, E. P. J.\& Lorimer, D. R., 1996, MNRAS, 283, L37
\bibitem{}
Wanajo S., 2013, ApJ, 770, L22
\bibitem{}
Wanajo S., Kajino T., Mathews G. J., Otsuki K., 2001, ApJ, 554, 578
\bibitem{}
Wheeler J. C., Cowan J. J., Hillebrandt W., 1998, ApJ, 493, L101
\bibitem{} Winteler, C., K\"appeli, R., Perego, A., Arcones, A., Vasset, N., Nishimura, N., Liebendörfer, M., Thielemann, F.-K.,
2012, ApJ, 750, L22
\bibitem{} Woosley, S.E. \& Weaver, T.A., 1995, ApJS, 101, 181
\bibitem{}
Woosley S. E., Wilson J. R., Mathews G. J., Hoffman R. D., Meyer B. S., 1994, 
ApJ, 433, 229
\end{thebibliography}
\end{document}